\documentclass[letterpaper, 10pt,conference]{ieeeconf}
\IEEEoverridecommandlockouts
\usepackage{amsmath,amssymb,amsfonts}
\usepackage{algorithmic}
\usepackage{graphicx}
\usepackage{textcomp}
\usepackage{xcolor}
\usepackage{url}

\usepackage{tikz}

\newtheorem{problem}{Problem}

\newtheorem{remark}{Remark}

\title{\Large \bf Electrical Fault Localisation Over a Distributed Parameter Transmission Line}

\author{Daniel Selvaratnam$^1$, Amritam Das$^2$, and Henrik Sandberg$^1$
	\thanks{$^{1}$Division of Decision and Control Systems, EECS, KTH Royal Institute of Technology, SE-100 44 Stockholm, Sweden. Email:
		{\tt\small \{selv, hsan\}@kth.se}. Supported in part by the Swedish Energy Agency and ERA-Net Smart Energy Systems (project RESili8, grant agreement No~883973) and the Swedish Research Council (Grant~2016-00861).
	}%
	\thanks{$^{2}$ Control Systems Group, EE Dept., Eindhoven University of Technology,
		P.O. Box 513, 5600 MB Eindhoven, The Netherlands. Email:
		{\tt\small am.das@tue.nl}}%
}

\newcommand{\R}{\mathbb{R}}

\newcommand{\m}{\mathrm{m}}

\DeclareMathOperator*{\argmin}{arg\,min}
\renewcommand\d{\mathop{}\!\mathrm{d}}

\begin{document}

\maketitle

\begin{abstract}
Motivated by the need to localise faults along electrical power lines, this paper adopts a frequency-domain approach to parameter estimation for an infinite-dimensional linear dynamical system with one spatial variable. Since the time of the fault is unknown, and voltages and currents are measured at only one end of the line, distance information must be extracted from the post-fault transients. To properly account for high-frequency transient behaviour, the line dynamics is modelled directly by the Telegrapher’s equation, rather than the more commonly used lumped-parameter approximations. First, the governing equations are non-dimensionalised to avoid ill-conditioning. A closed-form expression for the transfer function is then derived. Finally, nonlinear least-squares optimisation is employed to search for the fault location. Requirements on fault bandwidth, sensor bandwidth and simulation time-step are also presented. The result is a novel end-to-end algorithm for data generation and fault localisation, the effectiveness of which is demonstrated via simulation.
\end{abstract}

\section{Introduction}

Electrical faults in high-voltage transmission lines and medium-voltage distribution lines risk interrupting the power supply to large numbers of customers. Line-to-earth fault currents may, for example, result from interactions between power lines and growing tree branches, or from the gradual decay in cable insulation. Often there are early signs of such developing fault conditions, but the resulting earth currents have short duration and are not large enough to trigger the regular protection relays. These fault conditions can therefore go unnoticed until they cause a major power outage. However, with access to high-resolution, high-frequency measurements of voltage and current, it is possible to detect and estimate the location of developing faults at an early stage~\cite{habib_fault_2022}. Such detection systems have many advantages: The line/cable can be repaired before any customer is affected by an outage, and the repair can be scheduled ahead of time. A reasonable estimate of the distance to the fault can also simplify the repair process, since lines can easily span several kilometers. All these factors reduce the cost to both system operator and customer. 

There are four broad classes of methods to estimate fault location: Steady-state methods~\cite{hanninen_earth_2002,altonen_novel_2013}, transient-based methods~\cite{druml_new_2021,druml_results_2022}, travelling-wave methods~\cite{borghetti_integrated_2010,liang_fault_2015,el-hami_new_1992}, and data-driven methods~\cite{xing_physics-informed_2023,shakiba_real-time_2022}. Helpful overviews are provided in \cite{habib_fault_2022,parmar_fault_2015}.
 Many of these rely on measurements from at least two points on the line. Data-driven methods depend on large datasets for training, which may not be available. The steady-state and transient methods are typically based on lumped parameter models. In this work, we adopt a distributed parameter model of the transmission/distribution line in order to extract information from the high-frequency fault transients. This allows us to localise the fault with sensors at only one end of the line. Our work therefore relates most closely to the traveling wave methods~\cite{borghetti_integrated_2010,liang_fault_2015,el-hami_new_1992}, but differs in the following aspects. We perform nonlinear least-squares in the frequency domain, which has not been previously attempted for this problem. Information is thereby extracted from all frequencies within the sensor bandwidth, rather than only specific modes (c.f. \cite{liang_fault_2015}).  
In contrast to~\cite{borghetti_integrated_2010,el-hami_new_1992}, our approach exploits knowledge of the line dynamics by formulating it as an infinite-dimensional linear control system, with the fault as input and sensor measurements as output. 
We then derive the transfer function between input and output, without making finite-dimensional approximations, and use it to predict the output spectrum. A novel expression for the transfer function is obtained, in terms of the matrix exponential. Though non-rational, this presents no impediment to its use in least-squares estimation.

Our primary contribution is to demonstrate the potential of frequency-domain least-squares for electrical fault localisation, and identify the conditions under which it is effective. Since this is a preliminary work, we treat the problem in an idealised setting (see Remark \ref{rem:fidelity}). Noise is not considered, and the fault dynamics is approximated via a boundary condition. Based on the frequency response of the system, we propose guidelines for selecting sensor bandwidths, simulation time-steps, and modelling fault profiles, to facilitate the localisation. In particular, we observe that the fault profile must exceed a minimum bandwidth to make localisation possible. The simulation results show that, if the guidelines are followed, then even for bandlimited sensors at a single end of the line, the output spectrum contains sufficient information to localise the fault. Since least-squares is shown to be effective under idealised conditions, future work will build on it to accommodate more realistic assumptions. 

A possible reason for the surprising lack of other frequency-domain least-squares approaches in the literature is that testing of the algorithm requires the numerical solution of a hyperbolic PDE, in which information travels at the speed of light. We first non-dimensionalise the Telegrapher's equation to avoid ill-conditioning. This helps to preserve numerical stability both in simulation, and when evaluating the transfer function matrix exponential at different frequencies. We emphasise the former is only for testing --- the proposed estimator does not require the solution of a PDE. The non-dimensionalisation procedure is described in Section~\ref{sec:prob}, along with our problem formulation and modelling of the transmission line as a linear control system. In Section~\ref{sec:methodology}, the transfer function is derived, and the least-squares problem formulated. Section~\ref{sec:numerics} treats a numerical example as a case study to highlight important issues and present the design guidelines. The cost function is also plotted, and a global solution obtained. Concluding remarks are then made in Section \ref{sec:conclusion}, and directions identified for future work. 
\section{Problem Formulation}
\label{sec:prob}
\subsection{Modelling}
\begin{figure}[h]
	\begin{tikzpicture}
		\node[rectangle,draw] (sensor) at (0,0) {Sensor};
		\node[rectangle,draw] (fault) at (6,0) {Fault};
		\node[] (u) at (6,1) {$u(t - t_f)$};
		\node[] (y) at (0,1) {$y(t)$};
		\draw (sensor) -- (fault) node[midway, above]{$\begin{matrix} v(t,x)\\i(t,x) \end{matrix}$}; 
		\draw[thick, ->] (0,-1) -- (1,-1) node[anchor= west] {$x$};
		\fill (3,0) circle (2pt);
		\draw[->] (sensor) -- (y);
		\draw[->] (u) -- (fault);
		\draw[thick] (0,-1.1) -- (0,-0.9);
		\draw (fault) -- (7,0);
		\node[] at (6,-0.5) {$x = \ell$};
		\node[] at (0,-0.5) {$x = 0$};
	\end{tikzpicture}
\caption{Distribution line diagram \label{fig:diagram}}
\end{figure}
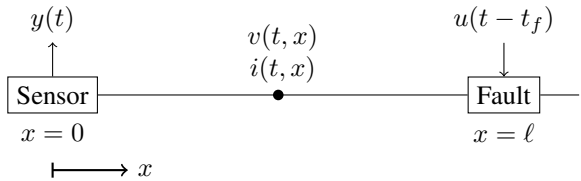
Consider Figure \ref{fig:diagram}. Let $v(t,x)$ denote the voltage at time $t \geq 0$ and position $x  \geq 0$ along an electrical distribution line, and $i(t,x)$ the corresponding current. Sensors measure the voltage and current at position $x = 0$. Suppose a fault occurs at position $\ell > 0$ along the line. Restricting attention to the segment of the line between fault and sensors, we assume both $v:[0,\infty) \times [0,\ell] \to \R $ and $i:[0,\infty) \times [0,\ell] \to \R $ are continuously differentiable maps. The spatio-temporal relationship between current and voltage is then modelled by the Telegrapher's equation
\begin{align}
\label{eq:telegraphers}
\begin{bmatrix}
	C & 0 \\ 0 & L
	\end{bmatrix}
 \begin{bmatrix}
   \frac{\partial v}{\partial t}\\
    \frac{\partial i}{\partial t}
\end{bmatrix} = \begin{bmatrix}
    0 & -1\\
    -1 & 0
\end{bmatrix}\begin{bmatrix}
   \frac{\partial v}{\partial x}\\
    \frac{\partial i}{\partial x}
\end{bmatrix} - \begin{bmatrix}
    G & 0\\
    0 & R
\end{bmatrix} \begin{bmatrix}
   v\\
   i
\end{bmatrix},
\end{align}
where $R, L,  C, G \geq 0$ are constants denoting, respectively, the distributed line resistance, inductance, capacitance, and conductance, per unit length. We assume, in particular, that $R,L,C > 0$. The goal is to estimate the fault location $\ell$ from the sensor measurements $v(t,0)$ and $i(t,0$).

\subsection{Nondimensionalisation} \label{sec:nondimen}
Typical line parameter values in SI units (c.f. \cite{glover_power_2022,thorslund_swedish_2017}), can lead to highly ill-conditioned coefficient matrices in \eqref{eq:telegraphers}, which pose severe numerical difficulties for both simulation and estimation. To circumvent this, we work with a partially non-dimensional form of the Telegrapher's equation, derived below. Currents and voltages are normalised by given base values $I_0$ and $V_0$, respectively. The resulting normalised values correspond to the per-unit system, common in the power systems literature~\cite[Chapter 3.3]{glover_power_2022}.  Distance is normalised by fault location $\ell$, which now appears as an unknown parameter in the dynamics. Time is not re-scaled. 
Accordingly, we define $z:[0,\infty) \times [0,1] \to \R^2$ as

$$ z(t,\xi) = \begin{bmatrix}
	z_1(t,\xi) \\ z_2(t,\xi)
\end{bmatrix} := \begin{bmatrix}
	\frac{v(t,\ell \xi)}{V_0} \\ \frac{i(t,\ell \xi)}{I_0}
\end{bmatrix}, $$
where $z_1 = \frac{v}{V_0}$, $z_2 = \frac{i}{I_0}$, and $\xi = \frac{x}{\ell}$, are non-dimensional voltages, currents, and positions, respectively. Our choice of $V_0$ and $I_0$ defines a base resistance $R_0 := \frac{V_0}{I_0}$. 
Under this change of variables, \eqref{eq:telegraphers} takes the form
\begin{equation} \ell E \frac{ \partial z}{\partial t} = \Gamma \frac{ \partial z}{\partial \xi} - \ell Fz, \label{eq:dynamics} \end{equation}
where \begin{align}
 	&E:= \begin{bmatrix}
 		CR_0 & 0 \\ 0 & \frac{L }{R_0} 
 	\end{bmatrix}, &&F := \begin{bmatrix}
 	GR_0 & 0 \\ 0 & \frac{R}{R_0}
 \end{bmatrix}, \label{eq:EF} \\ &\Gamma: = \begin{bmatrix}
 0 & -1\\
 -1 & 0
\end{bmatrix}. \label{eq:Gamma}
 \end{align}
\begin{remark}
	The matrix $\Gamma$ is involutory; that is, $\Gamma = \Gamma^{-1}$. 
	\end{remark}
\begin{remark} Since typically $C \ll L$ and $G \ll R$, a comparison of \eqref{eq:EF} with \eqref{eq:telegraphers} reveals that non-dimensionalisation improves the problem conditioning by a factor of $R_0^2$. 
	\end{remark} 
\subsection{Initial and boundary conditions}
The fault input and sensor outputs can now be imposed as boundary conditions on \eqref{eq:dynamics}. Sensors measure both voltage and current over time at $\xi=0$, yielding an output signal $y:[0,\infty) \to \R^2$ given by
\begin{equation}
\forall t \geq 0,\ y(t) = z(t,0). \label{eq:outputBoundary}
\end{equation}
We have chosen here to remain in dimensionless units. Note, nondimensionalisation of the physical sensor outputs only requires the known base quantities $I_0$ and $V_0$. 

By definition, the fault occurs at non-dimensional distance $\xi = 1$. Suppose it starts at an unknown time $t_f > 0$. The initial voltage and current distributions along the line are assumed known. Since \eqref{eq:dynamics} is a system of linear PDEs, the superposition principle applies, and therefore the zero initial condition
\begin{equation}
 \forall \xi \in [0,1],\    z(0,\xi) = 0 \label{eq:ICs}
\end{equation}
can be imposed without loss of generality. Assuming, in addition, a constant ratio between voltage and current at the fault location,
$$ \forall t \geq 0,\ v(t,\ell) = r i(t,\ell),$$
where $r \geq 0$ is the fault resistance. 
After non-dimensionalisation, the fault can then be modelled via the boundary condition
\begin{equation}
 \forall t \geq 0,\   z(t,1) = B u(t - t_f), \label{eq:inputBoundary}
\end{equation}
where $$B := \begin{bmatrix}
	\frac{r}{R_0} \\ 1
\end{bmatrix},$$ and $u:\R \to \R$ is a known scalar disturbance profile that describes the temporal behaviour of the fault.
Together, \eqref{eq:dynamics} -- \eqref{eq:inputBoundary} constitute an infinite-dimensional linear system with input $u$ and output $y$. 
\begin{problem}[Fault localisation] \label{prob:faultLoc}
Given the constants $R,L,C,G>0$, system model \eqref{eq:dynamics} -- \eqref{eq:inputBoundary}, input profile $u$, and output signal $y$, estimate $\ell$ without knowing $t_f$. 
\end{problem}
\begin{remark}[Model fidelity] \label{rem:fidelity}
The model \eqref{eq:dynamics} -- \eqref{eq:inputBoundary} ignores some phenomena present in a real transmission/distribution system. It can be viewed as describing a fault that is much closer to the measurement point than any other node on the line. Noise is neglected. Branches in the line, and the presence of loads on either side of the fault are not considered. The input boundary condition \eqref{eq:inputBoundary} does not allow for dynamic interactions between the fault and the line currents at the fault location itself. For example, it does not account for reflections returning from the unmeasured end of the line. A complex impedance between fault voltage and current is to be expected, but \eqref{eq:inputBoundary} only models the resistive part. Despite these simplifications, the Telegrapher's equation suffices to show how information about the fault location is preserved at different frequencies across the sensor output spectrum. This leads to the design guidelines of Section \ref{sec:numerics}. 
\end{remark}
\section{Methodology} \label{sec:methodology}
\subsection{Computing Input-Output Properties}
Since \eqref{eq:dynamics} -- \eqref{eq:inputBoundary} represents a linear system, we exploit the notion of a \emph{transfer function} to describe the relationship between the input \eqref{eq:inputBoundary} and the output \eqref{eq:outputBoundary} according to 
\begin{align}
\label{eq:IO}
    Y(s) = H(s; \ell) e^{-t_fs}U(s).
\end{align}
where $Y(s)$ and $U(s)$ are the Laplace transforms of $y(t)$ and $u(t)$, respectively. The mapping $s \mapsto H(s; \ell)$ is the $2 \times 1$ system transfer function parameterized by fault location $\ell$.

To derive $H$, we take the usual Laplace transform of \eqref{eq:telegraphers}. Since the initial condition is zero, we obtain the following Ordinary Differential Equation (ODE)
\begin{align*}
     \Gamma \frac{ \partial Z}{\partial \xi}(s, \xi) = \ell E s Z(s, \xi) + \ell FZ(s, \xi),
\end{align*}
the solution of which is given by
\begin{align}
\label{sol_ode}
    Z(s, \xi) = e^{\Gamma (F+Es)\ell\xi} Z(s, 0).
\end{align}
Here, we use the fact that $\Gamma$ is involutory. Since \eqref{sol_ode} is continuous and well-defined for all $\xi \in [0, 1]$, substituting $\xi = 1$ yields 
\begin{align}
\label{sol_ode1}
    Z(s, 1) = e^{\Gamma (F+Es)\ell} Z(s, 0). 
\end{align}

Using the Laplace transform of \eqref{eq:inputBoundary} and \eqref{eq:outputBoundary}, and substituting them into \eqref{sol_ode1} yields the transfer function as follows:
\begin{align}
    Y(s) = \underbrace{e^{-\Gamma (F+Es)\ell}B}_{=H(s;\ell)} e^{-t_fs}U(s). \label{eq:transferFunction}
\end{align}

\subsection{Proposed least-squares estimation in Frequency domain}
Expressing \eqref{eq:IO} component-wise,
\begin{equation*}
	\begin{bmatrix}
		Y_1(s) \\ Y_2(s)
	\end{bmatrix} = e^{-t_fs}\begin{bmatrix}
		H_1(s;\ell) \\ H_2(s;\ell)
	\end{bmatrix} U(s),
\end{equation*}
and evaluating the magnitude of each component along the imaginary axis,
\begin{equation}
	\begin{bmatrix}
		|Y_1(j \omega)| \\ |Y_2(j \omega)|
	\end{bmatrix} = \begin{bmatrix}
		|H_1(j \omega;\ell)| \\ |H_2(j \omega; \ell)|
	\end{bmatrix} |U(j \omega)|. \label{eq:mags}
\end{equation}
This equation relates the output voltage and current spectra to the fault spectrum. Note that taking magnitudes has removed dependence on the unknown fault time $t_f$. Acknowledging that all sensors have only a finite bandwidth $\omega_b > 0$ over which their output spectra can be accurately computed, the least-squares estimate
\begin{equation} \hat{\ell} = \argmin_{\ell \geq 0} \int_0^{\omega_b} \left\| \begin{bmatrix}
		|Y_1(j \omega)| \\ |Y_2(j \omega)|
	\end{bmatrix} - \begin{bmatrix}
		|H_1(j \omega;\ell) U(j \omega)| \\ |H_2(j \omega; \ell)U(j \omega)|
	\end{bmatrix} \right\|^2 \d \omega \label{eq:LS} \end{equation}
offers a solution to Problem \ref{prob:faultLoc}. The bandwidth $\omega_b$ plays an important role in the estimation problem, which is discussed in Section \ref{sec:BW}. 
In practice, only a sampled version of the output $y$ is available. Assuming a fixed sampling interval $T_s>0$, the output spectrum can be computed at discrete frequencies $\omega_k:= \frac{\pi k}{N T_s}$ via a $2N$-point FFT, and then a least-squares estimate obtained by minimising the cost $J:\R \to [0,\infty)$,
\begin{equation} J(\ell):= \sum_{k=0}^N  \left\| \begin{bmatrix}
		|Y_1(j \omega_k)| \\ |Y_2(j \omega_k)|
	\end{bmatrix} - \begin{bmatrix}
		|H_1(j \omega_k;\ell) U(j \omega_k)| \\ |H_2(j \omega_k; \ell)U(j \omega_k)|
	\end{bmatrix} \right\|^2. \label{eq:costJ} \end{equation}
\section{Numerical Case Study}
\label{sec:numerics}
\subsection{Problem data}
\label{sec:data}
We now tackle a particular instance of Problem \ref{prob:faultLoc} involving a 220kV distribution line in the Swedish power grid. The line parameter values in Table \ref{tab:data} are derived from \cite[Table 4]{thorslund_swedish_2017}, after conversion into SI units.
\begin{table}[h]
\caption{Parameter values for fault localisation case study \label{tab:data}}
	\begin{tabular}{ |c|c|c|c| } 
		\hline
		Parameter & Symbol & Value & Units \\
		\hline
		Line resistance & $R$ & $ 5.3900 \times 10^{-5}$ & $\Omega/\m$ \\ 
		Line inductance & $L$ & $1.3114 \times 10^{-6}$ & $\mathrm{H}/\m$ \\ 
		Line capacitance & $C$ & $9.1001 \times 10^{-12}$ & $\mathrm{F}/\m$ \\ 
		Line admittance & $G$ & 0 & $\mathrm{S}/\m$ \\ 
		Base voltage & $V_0$ & $220 \times 10^3$ \ & V \\ 
		Base current & $I_0$ & 454.55 & A \\ 
		Fault resistance & $r$ & 5 & $\Omega$ \\ 
		Fault time & $t_f$ & 0.01 & s \\ 
		Fault distance & $\ell$ & 2000 & m \\ 
		\hline
	\end{tabular}
\end{table}
We now simulate the fault in Table \ref{tab:data}, and estimate its location using the least-squares procedure of Section \ref{sec:methodology}. This example serves as a case study to illustrate the main steps involved, highlight important issues, and develop guidelines for both design and simulation.
\subsection{Bandwidth considerations} \label{sec:BW}
Figure \ref{fig:transferFunction} plots the magnitude spectrum of the transfer function $H$ for the given problem data, at different values of $\ell$. It illustrates a general pattern that can be empirically observed: \begin{itemize}
	\item For each value of $\ell$, the frequency response of both components are flat until a critical frequency $\omega^\star$, at which they begin to change.
	\item The DC gains $h_1:=|H_1(0;\ell)|, h_2:= |H_2(0;\ell)|$ are approximately constant with $\ell$.
	\item The critical frequency $\omega^\star$ decreases with $\ell$. 
	\end{itemize}
Such behaviour has important practical ramifications for modelling, sensing, and simulation. 
\subsubsection{Fault bandwidth} \label{sec:faultBW}
In order to estimate fault location, the fault profile $u$ must contain frequencies higher than $\omega^\star$. To see this, suppose that $|U(j\omega)| \approx 0$ for $\omega \geq \omega_f$. It follows that $$|H_i(j\omega;\ell)U(j \omega)| \approx \begin{cases}
	h_i|U(j \omega)|,& \omega \in [0,\omega^\star] \\
	|H_i(j\omega;\ell)||U(j \omega)|,& \omega \in (\omega^\star,\omega_f)\\
	0,& \omega \geq \omega_f	
\end{cases}.$$
If the fault bandwidth $\omega_f \leq \omega^\star$, this removes all dependence on $\ell$ from the cost function $J(\ell)$ in \eqref{eq:costJ}, making the optimisation problem degenerate.  
\subsubsection{Sensor bandwidth} \label{sec:sensorBW}
The sensor bandwidth $\omega_b$ must also exceed $\omega^\star$ to capture $|Y(j \omega)|$ at the informative frequencies. Since $\omega^\star$ decreases with $\ell$, the further away a fault is, the more easily it can be localised (that is, localised with less expensive equipment). A system designer could first specify a minimum distance for reliable localisation, which determines $\omega^\star$, and then choose a sensor with bandwidth $\omega_b > \omega^\star$. 
\begin{figure}
	\includegraphics[width = \linewidth]{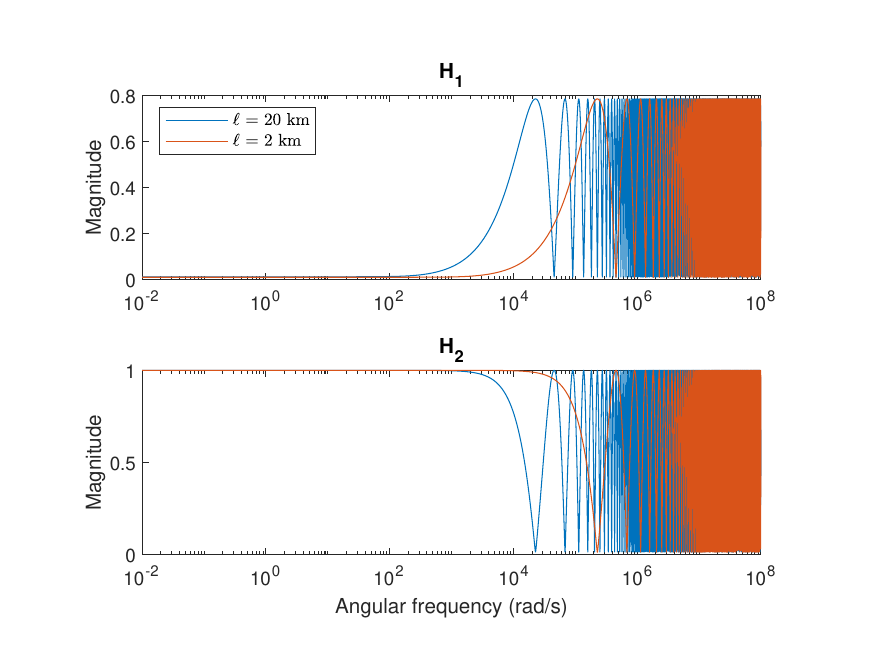}
	\caption{Magnitude spectrum of $H(s;\ell)$, given data in Table \ref{tab:data}, for different fault distance values. \label{fig:transferFunction}}
\end{figure}

\subsection{Fault modelling}
\begin{figure}
	\includegraphics[width = \linewidth]{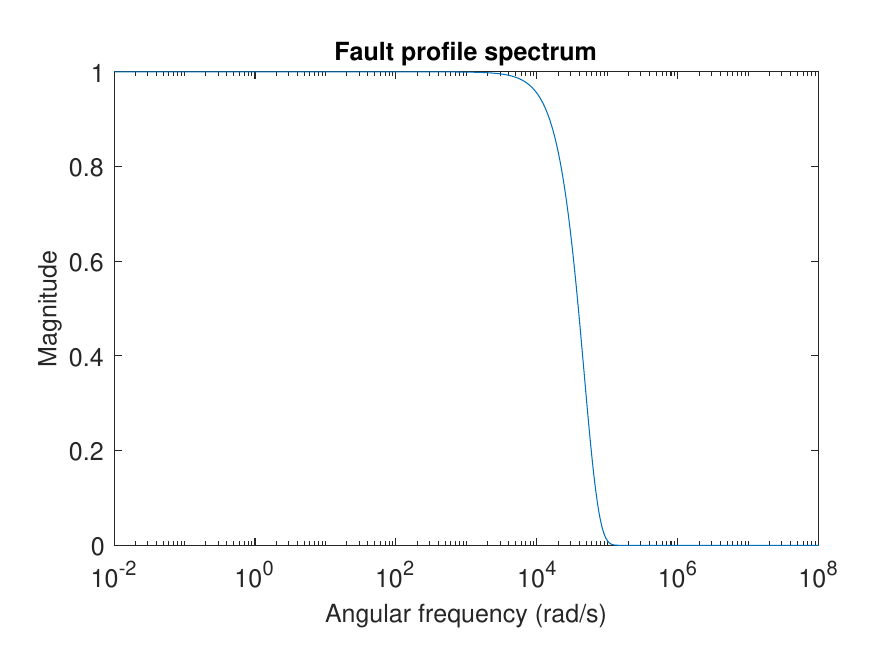}
	\caption{Magnitude spectrum of fault profile, for $\sigma = 3 \times 10^{-3}$. \label{fig:inputSpectrum}}
\end{figure}

Our proposed estimation scheme relies on an assumed fault profile $u$. Based on the discussion in Section \ref{sec:faultBW}, the ideal fault (from a localisation perspective) is an impulse, because it excites all frequencies equally. Such is also consistent with the sudden and fleeting nature of line-to-earth faults. However, the finite sensor bandwidth $\omega_b$ implies that all information beyond this frequency is lost. In simulation, numerical PDE solvers must resort to finite time-steps, so they are also unable to reproduce dynamics beyond a certain frequency. For simulation purposes, we therefore model the fault as a Gaussian pulse
$$ u(t) = \frac{1}{\sigma\sqrt{2 \pi}}e^{- \frac{t^2}{2 \sigma^2}},$$
which approaches the unit impulse as $\sigma \to 0$. Its Fourier Transform is a Gaussian pulse in the frequency domain,
$$ U(j \omega) = e^{-\frac{\omega^2 \sigma^2}{2}},$$ which is clearly bandlimited. 
Referring to Figure \ref{fig:transferFunction}, the critical frequency for a fault at $\ell = 2$ km is $\omega^\star \approx 10^4$ rad/s.  In accordance with Section \ref{sec:faultBW}, we choose $\sigma = 3.0349 \times 10^{-5}$ s to obtain a fault bandwidth of $\omega_f = 10^5\ \mathrm{rad/s} > \omega^\star$. The fault spectrum is plotted in Figure \ref{fig:inputSpectrum}. 
\begin{remark} \label{rem:faultMod}
A Gaussian fault profile is chosen here as an example of a bandlimited pulse, because PDE solvers and physical sensors cannot reproduce arbitrarily high frequencies. 
In reality, it is difficult to know the true current and voltage signals at the fault location, because sensors are never present exactly where a fault occurs. Different types of faults are possible~\cite[Chapters 8 \& 10]{glover_power_2022}, and they are expected to produce different profiles. Future work will explore the use of unknown input observers~\cite{ghanipoor_ultra_2022,chen_design_1996} to deal with uncertainty in the fault profile. In a nominally balanced 3-phase system, it may even be possible to estimate the fault current profile from the symmetrical zero sequence component.
\end{remark}
\subsection{Simulation of PDE dynamics}
\begin{figure}
 	\includegraphics[width = \linewidth]{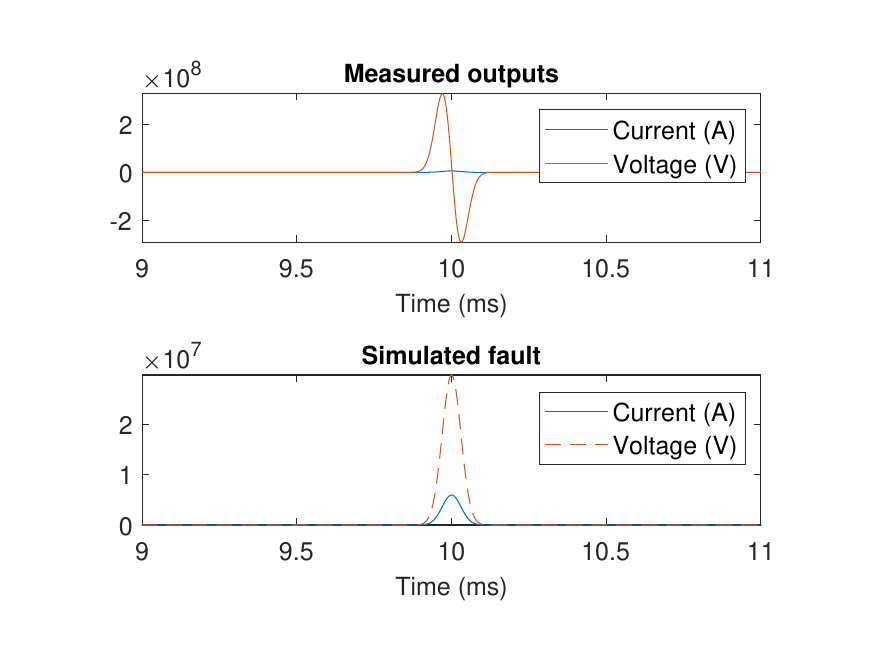}
 	\caption{Simulated current and voltage at fault location, together with corresponding sensor outputs. \label{fig:simOutput}}
 \end{figure}

A fixed simulation time-step of $T_s > 0$ effectively models a sensor with the same sampling period, which corresponds to a sensor bandwidth $\omega_b = \frac{\pi}{T_s}$ equal to the Nyquist frequency. In order to faithfully simulate the line dynamics over the frequency range $[\omega^\star, \omega_f]$, we must have $\omega_b \gg \omega_f$. 
For this case study, let $T_s = \frac{\pi}{10\omega_f}$, which corresponds to $\omega_b = 10 \omega_f$. To simulate the line dynamics \eqref{eq:dynamics}, we have used the PIETOOLS~\cite{shivakumar_pietools_2022} numerical PDE solver, which employs Petrov-Galerkin projection onto a polynomial basis, together with the backward-difference discretisation of temporal derivatives. The simulated sensor outputs are plotted in Figure \ref{fig:simOutput}. The non-dimensionalisation in Section \ref{sec:nondimen} is necessary for backward-difference to be numerically stable under this choice of time-step and parameter values, but it may not remain so for arbitrarily small time-steps. If numerical instability is encountered for the chosen time-step, other integration schemes may be required. In particular, \cite{mohanty_high-precision_2023,sharifi_numerical_2016,saadatmandi_numerical_2010,lakestani_numerical_2010,dehghan_numerical_2008} propose numerical methods that are specialised for the Telegrapher's equation.

For a real-world localisation problem, the voltage and current initial conditions would reflect the steady-state solution of the power line in response to sinusoidal forcing. However, due to linearity, the corresponding frequency component can simply be removed from the output spectrum before performing the minimisation. This step has been omitted for simplicity. Instead, the line has been simulated with zero initial conditions, as per \eqref{eq:ICs}.
 
\subsection{Estimation}
If the preceding guidelines for fault bandwidth, sensor bandwidth and simulation time-step are followed, the least-squares procedure itself is straightforward. Evaluation of the cost function \eqref{eq:costJ} requires computation of the matrix exponential in \eqref{eq:transferFunction}, which is numerically unstable without the non-dimensionalisation in Section \ref{sec:nondimen}.

\begin{figure}
	\includegraphics[width = \linewidth]{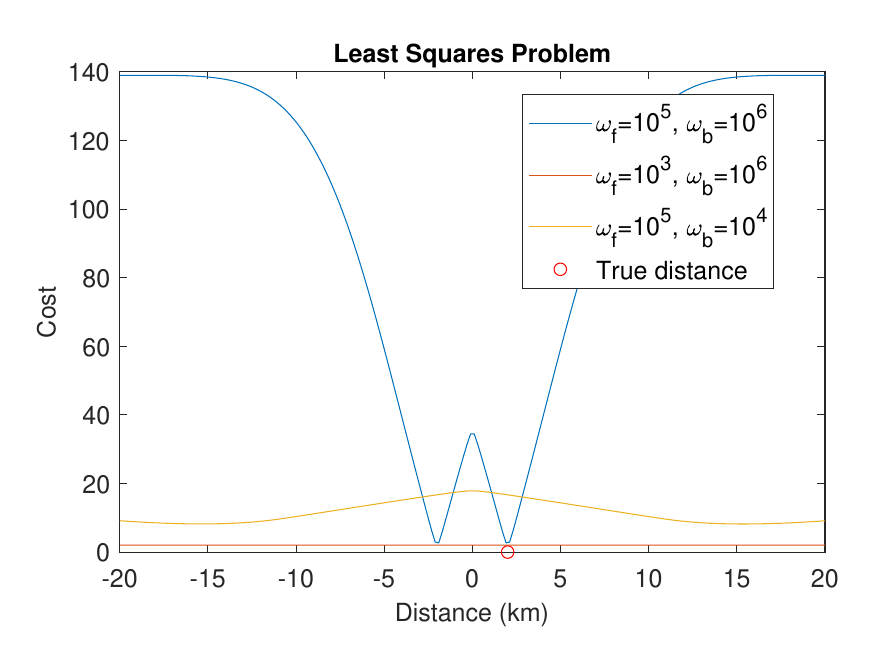}
	\caption{Cost $J(\ell)$ as a function of $\ell$, under different fault and sensor bandwidths. Blue plot: $\omega_f = 10 \omega^\star,\ \omega_b = 100 \omega^\star $. Red plot: $\omega_f = 0.1 \omega^\star,\ \omega_b = 100 \omega^\star $. Yellow plot: $\omega_f = 10 \omega^\star,\ \omega_b = \omega^\star $. Frequencies are in rad/s. \label{fig:costFunc}}
\end{figure}

To illustrate the effect of fault bandwidth and simulation time-step on estimation quality, Figure \ref{fig:costFunc} plots the cost function for different values of $\omega_f$ and $\omega_b = \frac{\pi}{T_s}$. For the blue line, the guidelines are followed, with bandwidth values as previously stated: $\omega_b = 10 \omega_f$ and $\omega_f = 10 \omega^\star $. For the red line, the fault bandwidth $\omega_f$ is reduced to a tenth of the critical frequency $\omega^\star$, and for the yellow line, the sensor bandwidth $\omega_b$ reduced to a tenth of the fault bandwidth. The latter are almost completely uninformative.

Focusing on the blue line, its shape suggests that gradient descent will converge to one of two global minima, for any non-zero initial condition. The observed symmetry means that unconstrained optimisation can be employed, if the sign of the result is ignored. For this instance of Problem \ref{prob:faultLoc}, given an initial guess of 1~m (i.e., a 1.999~km initial error), \textsc{Matlab}'s \texttt{fminunc} returns a location estimate within 4.21~s with a 5.39~cm error. A standard HP Elitebook was used, with Intel CORE i7 vPRO processor and 32~GB RAM, running \textsc{Matlab} R2022b on Windows~10.
\section{Conclusion} \label{sec:conclusion}
A fault localisation algorithm for electrical distribution lines is presented. 
A distributed parameter model for the line, the Telegrapher's equation, is adopted. This yields an infinite-dimensional dynamical system, with the fault as an input applied at one end of the domain, and sensor measurements as an output from the opposite end. First, the transfer function between input and output is derived. The estimator then performs nonlinear least-squares between the predicted and measured output spectra, based on an assumed fault input profile. The resulting cost function does not depend on the unknown fault time. A numerical case study is then presented. Based on the derived frequency response of the system, guidelines for the selection of sensor bandwidths, fault pulse bandwidths, and simulation sampling times are offered. 

The numerical results obtained suggest several directions for both theoretical and practical development.
First, to establish analytically the three empirical observations made in Section~\ref{sec:BW}. Second, identifying mathematical properties of the cost function that guarantee convergence to global minima. Third, relaxing the modelling assumptions in Remarks~\ref{rem:fidelity} and \ref{rem:faultMod}. This involves more detailed modelling of the fault boundary condition, possibly resulting in a coupled ODE-PDE system, and the use of unknown input observers~\cite{ghanipoor_ultra_2022} to accommodate uncertainty in the fault input dynamics. Fourth, making the algorithm robust to parameter uncertainty and sources of noise, which may require estimation of additional unknown fault or line parameters. On the practical side, testing of the algorithm on real-world distribution line data is in progress. Adapting the procedure for use on three-phase distribution lines is also an important step towards potential industrial adoption of the algorithm. 

\bibliographystyle{ieeetr}
\bibliography{references}

\end{document}